\documentclass[twocolumn]{aastex62}
\usepackage{bm}
\usepackage[T1]{fontenc} 
\usepackage{subfigure}

\hypersetup{linkcolor=red,citecolor=blue,filecolor=cyan,urlcolor=magenta}

\catcode`_ = \active
\newcommand_[1]{\sb{\mathrm{#1}}}

\shorttitle{triple-ridge structure of AGN jets}
\shortauthors{Ogihara et al.}

\begin{document}

\title{A MECHANISM FOR TRIPLE-RIDGE EMISSION STRUCTURE OF AGN JETS}
\author{Taiki Ogihara}
\affiliation{Astronomical Institute, Tohoku University, Sendai, 980-8578, Japan}
\author{Kazuya Takahashi}
\affiliation{Center for Gravitational Physics, Yukawa Institute for Theoretical Physics, Kyoto University, Kyoto, 606-8502, Japan}
\author{Kenji Toma}
\affiliation{Frontier Research Institute for Interdisciplinary Sciences, Tohoku University, Sendai, 980-8578, Japan}
\affiliation{Astronomical Institute, Tohoku University, Sendai, 980-8578, Japan}


\begin{abstract}
  Recent radio VLBI observations of the relativistic jet in M87 radio galaxy have shown a triple-ridge structure that consists of the conventional limb-brightened feature and a central narrow ridge.
  Motivated by these observations, we examine a steady axisymmetric force-free model of a jet driven by the central black hole (BH) with its electromagnetic structure being consistent with general relativistic magnetohydrodynamic simulations, and find that it can produce triple-ridge images even if we assume a simple Gaussian distribution of emitting electrons at the base of the jet.
  We show that the fluid velocity field associated with the electromagnetic field produces the central ridge component due to the relativistic beaming effect, while the limb-brightened feature arises due to strong magnetic field around the jet edge which also induces the electrons to be dense there.
  We also show that the computed image strongly depends on the electromagnetic field structure, viewing angle, and parameters related to the electrons' spatial distribution at the jet base.
  This study will help constraining the non-thermal electron injection mechanism of BH jets and be complementary to theoretical analyses of the upcoming data of Event Horizon Telescope.
\end{abstract}



\section{INTRODUCTION} \label{sec: introduction}

Relativistic collimated outflows (or jets) have been observed in active galactic nuclei (AGNs).
It is widely thought that they are formed in a system composed of a black hole (BH) at the center of a galaxy and surrounding plasma.
Their plausible formation mechanism is electromagnetic extraction of the rotational energy of the BH and/or its accretion disk \citep{Blandford1977,Blandford1982,Uchida1985,Lovelace1987,Meier2001,Komissarov2004,Beskin2010}.
The former is called the Blandford-Znajek (BZ) process, and the latter is called the Blandford-Payne (BP) process.
General relativistic magnetohydrodynamic (GRMHD) simulations show that the globally ordered magnetic field is realized only in the funnel region around the rotation axis, where relativistic jet appears to be formed via the BZ process, and the disk wind is non-relativistic \citep[e.g.][]{McKinney2004,McKinney2009,Tchekhovskoy2011,Sadowski2013,Nakamura2018}.
In recent studies, the radiative transfer calculations are performed based on the GRMHD simulation results \citep[e.g.][]{Moscibrodzka2007,Dolence2009,Broderick2010,Porth2011,Dexter2012,Moscibrodzka2016,Pu2016}.
These studies enable us to compare the numerical results to observations.
If the BZ process is observationally confirmed, the existence of ergosphere would be indirectly supported.
\citep{Komissarov2004,Toma2014,Toma2016,Kinoshita2018}.

Despite those sophisticated computations, there remains the so-called `mass-loading problem' for relativistic jets.
No particle is injected into the jet from the BH, of course, and the globally ordered magnetic field prevents the surrounding thermal plasma particles from diffusing into the jet.
The origin of particles in jets is still unclear.
Electron-positron pair creation by the ambient photons \citep{Levinson2011,Moscibrodzka2011} and by the high energy photons emitted by electrons accelerated in the MHD-violated region or gap \citep{Beskin1992,Hirotani1998,Levinson2011,Broderick2015,Hirotani2016,Ptitsyna2016,Levinson2017} have been discussed for particle injection mechanism.\footnote{Very recently physics of the gaps in BH magnetospheres has been studied with particle-in-cell simulations \citep{Levinson2018, Chen2018, Parfrey2018}.}
High-energy hadron injection from the surrounding plasma \citep{Toma2012, Kimura2014, Kimura2015} and magnetic reconnection induced by the accretion of fields with alternating polarity \citep{Parfrey2015} might also be relevant.
In the GRMHD simulations, these non-thermal processes are not taken into account. The density of the jet usually becomes very low and is replaced by a density floor. Although this treatment does not affect the electromagnetic field structure because the particle energy density is much smaller than the electromagnetic field one in the funnel region, the terminal Lorentz factor and the emission of the jet depend directly on the particle density distribution \citep[][hereafter, T18]{McKinney2006, Moscibrodzka2016, Jeter2018, Takahashi2018}.
The spatial distribution of emitting particles is still the serious ambiguity when one compares the simulation results with observations.

Radio VLBI observations can resolve AGN jets and have reported the limb-brightened structure in jets of M87 \citep{Kovalev2007,Walker2008,Hada2011,Hada2016,Mertens2016,Kim2018}, Mrk 501 \citep{Piner2008}, Mrk 421 \citep{Piner2010}, Cyg A \citep{Boccardi2016}, and 3C84 \citep{Nagai2014,Giovannini2018}.
The M87 galaxy's jet has been actively observed because of its proximity \citep[$D \simeq 16.7$ Mpc,][]{Mei2007,Blakeslee2009} and its length ($\sim 10$ kpc).
Its central BH mass $M_{BH}$ is estimated as $(3.3-6.6) \times 10^9 \, M_{\odot}$\citep{Macchetto1997, Gebhardt2009, Gebhardt2011, Walsh2013}, and then the angular size of the Schwarzschild radius $R_{S} = 2GM_{BH}/c^2$ is $\approx 3.9-7.8$ $\mu$as.
The BH shadow and the jet launching region of M87 with this size is expected to be resolved by Global VLBI observation project Event Horizon Telescope (EHT) \citep{Doeleman2012, Akiyama2017}.
The limb-brightened structure of the M87 jet is confirmed between $\sim 40-10^5 \, R_{S}$ from the center, which are de-projected lengths under the assumption of the viewing angle $\Theta = 15^{\circ}$ \citep[c.f.][]{Wang2009}.

Among many theoretical papers that compute the synchrotron images of MHD jets, only a few discussed the origin of the limb-brightened structure \citep{Zakamska2008, Gracia2009, Moscibrodzka2016}.
T18 showed limb-brightened images by using an analytic force-free model consistent with GRMHD simulation results.
They showed that toroidal velocity of the flow (and its effect on relativistic beaming) is relevant for interpreting the image structure and that a jet from a rapidly spinning BH is favored.
Slowly spinning BHs and Keplerian disks do not efficiently accelerate the flow, leading to the large toroidal velocities and highly asymmetric images which are not consistent with the observations of the M87 jet.
It was also shown that sufficient amount of the electrons has to be injected on the magnetic field lines apart from the jet axis.
This implies that the characteristic image structure can potentially constrain the electron spatial distribution and the mass-loading mechanism.

In this paper, we focus on the newly discovered characteristic feature: High sensitivity observation with VLBA + phased-VLA \citep{Hada2017}, recent analysis of VSOP data \citep{Asada2016}, and the stacked VLBA images \citep{Walker2018} have revealed that the M87 jet image has `triple-ridge structure', i.e., a narrow central ridge in addition to the conventional limbs. Hereafter we call the central component `inner-ridge' and the limbs `outer-ridges'.
This feature might indicate that two different formation mechanisms work in a jet unlike the GRMHD simulation results, e.g. the BZ process for the inner-ridge and the BP process for the outer-ridges \citep{Sobyanin2017}. However, we show that a BZ jet can solely produce the triple-ridge images by following the formulation in T18. Thus, two different formation mechanisms are not necessarily required. We find that the MHD flow velocity structure of the BZ jets produces distinct component near the axis due to the relativistic beaming effect, which corresponds to the inner-ridge, while the outer-ridges may arise mainly because of stronger magnetic field and denser electrons around the jet edge. Note that we will simply illustrate this novel idea on the triple-ridge emission structure, not trying to fit the observed data of the M87 jet.

This paper is organized as follows: We briefly introduce our model in Section~\ref{sec: method}, and show the computed triple-ridge image and its parameter dependence along with their physical reasons in Section~\ref{sec: results}. We discuss some detailed properties of the observed triple-ridge structure in Section~\ref{sec: discussion}, and summarize our findings in Section~\ref{sec: summary}.


\section{MODEL} \label{sec: method}

To produce the synchrotron jet image, we use an analytic model which is essentially the same as used in T18.
Here, we explain this model briefly.
Readers can refer to T18 as well as \citet{Broderick2009} for more details.
Section~\ref{subsec: Force-Free Jet Model} introduces the modeling of electromagnetic field, velocity field, and density field structure, and
Section~\ref{subsec: Non-thermal energy distribution and radiation} treats electrons' energy distribution and their radiation.
We set up our parameter values in Section~\ref{subsec: model parameters}.
The quantities with prime denote those measured in the fluid rest frame.


\subsection{Force-Free Jet Model} \label{subsec: Force-Free Jet Model}
We assume that the force-free condition is valid in the funnel region, where particle inertia and pressure are not important for dynamics, as shown in GRMHD simulations and implied from the current observational data of the M87 jet \citep{Kino2014, Kino2015}.
We further put the steady axisymmetric condition, and then the poloidal magnetic field and electric field measured in the lab frame are described as
\begin{equation}
	\bm{B}_{p} = \frac{1}{R} \nabla\Psi \times \bm{\hat{\phi}},
	\label{eq: Bp}
\end{equation}
\begin{equation}
	\bm{E} = - \frac{1}{c} \Omega_{F} \nabla \Psi = - \frac{R\Omega_{F}}{c}\bm{\hat{\phi}} \times \bm{B}.
\end{equation}
($R, \phi, z$) are the standard cylindrical coordinates with the $z$ axis set to be the jet axis. $\bm{\hat{\phi}}$ is the azimuthal unit vector.
$\Psi$ is the magnetic flux function and $\Omega_{F}$ corresponds to the angular velocity.
We use a parametrically controlled form of $\Psi$,
\begin{equation}
	\Psi = A r^{\nu} (1 \mp \cos \theta),
	\label{eq: Psi}
\end{equation}
where $(r, \theta, \phi)$ are the standard spherical coordinates. $A$ is a constant that normalizes the magnetic field strength, and $\nu$ controls the jet shape. The plus and minus signs stand for $z<0$ and $z>0$, respectively.
Note that we focus on the emission structure far from the central BH, where we can neglect the GR effects. For this form of $\Psi$, the toroidal magnetic field is approximately given by \citep{Tchekhovskoy2008}
\begin{equation}
  B_{\mathrm{\phi}} = \mp \frac{2 \Omega_{F} \Psi}{Rc}.
  \label{eq: Bphi}
\end{equation}
Equations~(\ref{eq: Psi}) and (\ref{eq: Bphi}) are not exact solutions of the Maxwell equations under steady axisymmetric condition (i.e., Grad-Shafranov equation), but they fit well to GRMHD simulation results \citep{Tchekhovskoy2010, Nakamura2018}.
$\Psi$ and $\Omega_{F}$ are quantities conserved along each field line and the electromagnetic field is described by their distribution.
The cases of $\nu=0$ and $1$ correspond to the radial and parabolic shapes of $\bm{B}_{p}$, respectively.
For $\nu=2$, the cylindrical structure is obtained in the far zone.
As explained in Section~\ref{sec: introduction}, only the model with the field lines penetrating the rapidly rotating BH can produce the nearly symmetric limb-brightened images as observed (T18). We focus on this case, in which $\Omega_{F}(\Psi)$ is given as $\Omega_{F}(\Psi) \approx 0.5 \;\Omega_{H}$, where $\Omega_{H} = ac/2r_{+}$, $a$ is a spin parameter of the BH, and $r_{+}$ is the horizon radius of the rotating BH.
We set $a=0.998$.

We may not define the fluid velocity in principle in the force-free model, but practically we can use the drift velocity \citep{Tchekhovskoy2008, Beskin2010},
\begin{equation}
  \bm{v} = c \frac{ \bm{E} \times \bm{B} }{B^2} = R\Omega_{{\mathrm F}} \bm{\hat{\phi}} - R\Omega_{\mathrm F}\frac{B_{\phi}}{B^2} \bm{B}.
  \label{eq: drift velocity}
\end{equation}
This is consistent with the velocity in the Poynting-dominated cold ideal MHD formalism at the zone where $R\Omega_{F} \gg c$ (see Eqs. 45 and 46 of T18).
For $R\Omega_{F} \ll c$, the drift velocity has $v_\phi \sim R\Omega_{F} \gg v_{p}$ (Eqs. 42 and 43 of T18) for the electromagnetic field that we consider (as discussed later in Section~\ref{subsec: triple-ridge structure}), which is also consistent with the cold ideal MHD velocity. Cold ideal GRMHD calculations also show $v_\phi \gg v_{p}$ within the outer light cylinder \citep{Pu2015}, which supports our treatment of the fluid velocity.
We do not treat the case that the jet particles are injected along the magnetic field lines with a large Lorentz factor at the outflow base.
We briefly discuss this case in Section~\ref{subsec: inner-ridge property}.

We give the density distribution by the sourceless equation of continuity,
\begin{equation}
  \nabla \cdot (n \bm{v} ) = 0,
\end{equation}
where $n$ is the particle number density measured in the lab frame.
This equation combined with Equation~(\ref{eq: drift velocity}) leads to another conserved quantity,
\begin{equation}
  \frac{n}{B^2} = \mathrm{const. \, along \, each \, field \, line}.
  \label{eq:nb2const}
\end{equation}
When we give the density value at a certain point on each field line, we can obtain the density distribution everywhere.
We note that Equation~(\ref{eq:nb2const}) is consistent with $n v_p/B_p$ = const., that is the mass flux per unit magnetic flux conserved in the cold ideal MHD formalism. Indeed, one can obtain Equation~(\ref{eq:nb2const}) by substituting Equation~(\ref{eq: drift velocity}) to $n v_p/B_p$ = const.


\begin{table*}[htb!]
  \centering
  \caption{Model parameters}
  \begin{tabular}{llr}
    \hline
    Quality & symbol & value \\
    \hline
    \hline
    Mass of the BH & $M_{BH}$ & $6.2 \times 10^9M_{\odot}$ \\
    Dimensionless Kerr parameter of the BH & $a$ & 0.998 \\
    Rotational frequency of the magnetic field & $\Omega_{F}$ & $0.5 \Omega_{H}$\\
    Viewing angle & $\Theta$ & $15^{\circ}$ \\
    Jet Shape (Equation~\ref{eq: Psi}) & $\nu$ & 0.75 \\
    Radius where $n$ peaks at $z=\pm z_{1}$ & $R_{p}$ & 0 \\
    Height of the plane where $n$ is given by Equation~(\ref{eq:n}) & $z_{1}$ &  $5R_{S}$ \\
    Width of $n$ distribution (Equation~\ref{eq:n}) & $\Delta$ & $2R_{S}$ \\
    Energy spectral index of the non-thermal electrons & $p$ & 1.1 \\
    Minimal Lorentz factor of the non-thermal electrons & $\gamma'_{m}$ & 100 \\
    Luminosity distance to the jet & $D$ & 16.7 Mpc \\
    Inclination between the jet axis and the major axis of the beam kernel & & $16^{\circ}$ \\
    Beam size & & $1.14$ mas $\times$ $0.55$ mas \\
    \hline
  \end{tabular}
  \label{tab:parameters}
\end{table*}

\subsection{Non-thermal Electron Energy Distribution and Radiation} \label{subsec: Non-thermal energy distribution and radiation}
We set the fluid density at $z=z_1$ just for simplicity by
\begin{equation}
  n (R,z_1) = n_{0} \exp \left( - \frac{(R - R_{p})^2}{2 \Delta^2} \right),
  \label{eq:n}
\end{equation}
where $z_1, R_{p}, \Delta$ are free parameters.
We assume that a constant fraction of the electrons are non-thermal\footnote{Some other papers assume that the non-thermal electron density is related with the magnetic energy density \citep{Broderick2010, Porth2011, Dexter2012}.} and their energy distribution is described as
\begin{equation}
  f (\gamma') \propto \left\{
  \begin{array}{ll}
    \gamma'^{-p} & \mathrm{for}  \ \gamma' > \gamma'_{m} \\
    0 & \mathrm{for} \ \gamma' < \gamma'_{m},
  \end{array}
  \right.
  \label{eq: energy dist.}
\end{equation}
where $\gamma'$ is the Lorentz factor of electrons measured in the fluid frame, $\gamma'_{m}$ is the minimum value of $\gamma'$, and $p$ is the power-law index. $p$ and $\gamma'_{m}$ are set to 1.1 and 100 respectively throughout this paper.
We set the non-thermal electron density as zero along the field lines which do not penetrate the horizon, i.e., for $\Psi > \Psi(r=r_{+},\theta=\pi/2)$.

The synchrotron emissivity in the fluid frame $j_{\omega'}'$ is written by \citep{Rybicki1986}
\begin{eqnarray}
  j'_{\omega'} ({\hat{\bm n}}') =
  \frac{\sqrt{3} (p-1) e^3 n' B' \sin \alpha' }{8\pi^2 {\gamma'_m}^{1-p} m_e c^2 (p+1)} \left( \frac{m_e c \omega'}{3e B' \sin \alpha'} \right)^{ -\frac{p-1}{2} } \nonumber \\
  \times {\bar \Gamma} \left( \frac{p}{4} + \frac{19}{12} \right) {\bar \Gamma} \left( \frac{p}{4} - \frac{1}{12} \right), \nonumber \\
  \,
  \label{eq:jomega}
\end{eqnarray}
where $e, m_e, $ and ${\bar \Gamma}(x)$ are the elementary charge, the mass of electron, and the gamma function, respectively.
$B' \equiv |{\bm B'}|$, and $\alpha'$ is the pitch angle of the electrons defined by $\cos \alpha' = ({\hat {\bm n}}' \cdot {\bm B'})/|{\bm B'}|$, where ${\hat {\bm n}}'$ is the unit vector directing toward the observer in the fluid frame.
Then the emissivity at each point in the jet that is received by the observer is given by \citep{Rybicki1986,Shibata2003}
\begin{equation}
  j_{\omega} (\hat{\bm n}) = \frac{1}{\Gamma^2 (1-\beta\mu)^3} j_{\omega'}' ({\hat{\bm n}}') ,
  \label{eq: Lorentz trans.}
\end{equation}
where $\beta$ is the bulk speed normalized by the speed of light, and $\Gamma$ is its corresponding Lorentz factor. $\mu$ is cosine of the angle between the direction of the bulk flow and the direction of the line of sight in the lab frame.

The jet is optically thin for synchrotron radiation in the region where the triple-ridge structure is observed ($\sim 10 - 30$ mas from the radio core) and at the wavebands of those observations \citep{Asada2016,Hada2017}.
The observed intensity is then calculated by
\begin{equation}
  I_{\omega} (X,Y) = \int j_{\omega}(\hat{\bm n}, X, Y, Z) dZ ,
  \label{eq: intensity}
\end{equation}
where $dZ$ is the line element along the line of sight. $(X,Y)$ are the coordinates of the sky with the $Y$ axis being the projected jet axis. The viewing angle $\Theta$ is defined as the angle between the line of sight ($Z$ axis) and the jet axis ($z$ axis).
We first obtain the intensity map on the $X-Y$ plane by the computation of Equation~(\ref{eq: intensity})  with the spatial resolution of $3 R_{S}$, and obtain the simulated VLBI image after the convolution with a Gaussian beam kernel.


\begin{figure*}[htb!]
  \centering
  \includegraphics[clip,width=0.45\textwidth]{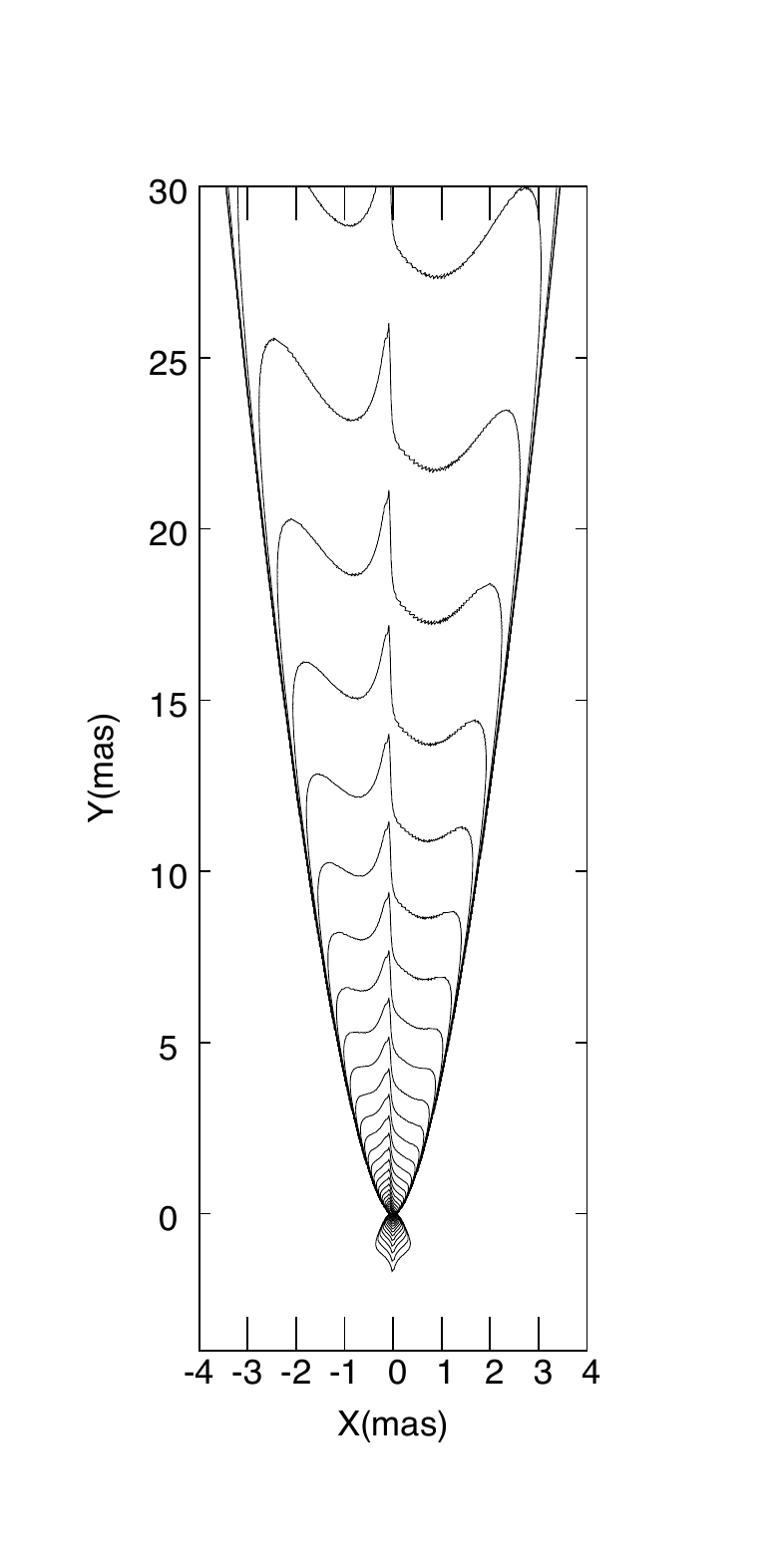}
  \includegraphics[clip,width=0.45\textwidth]{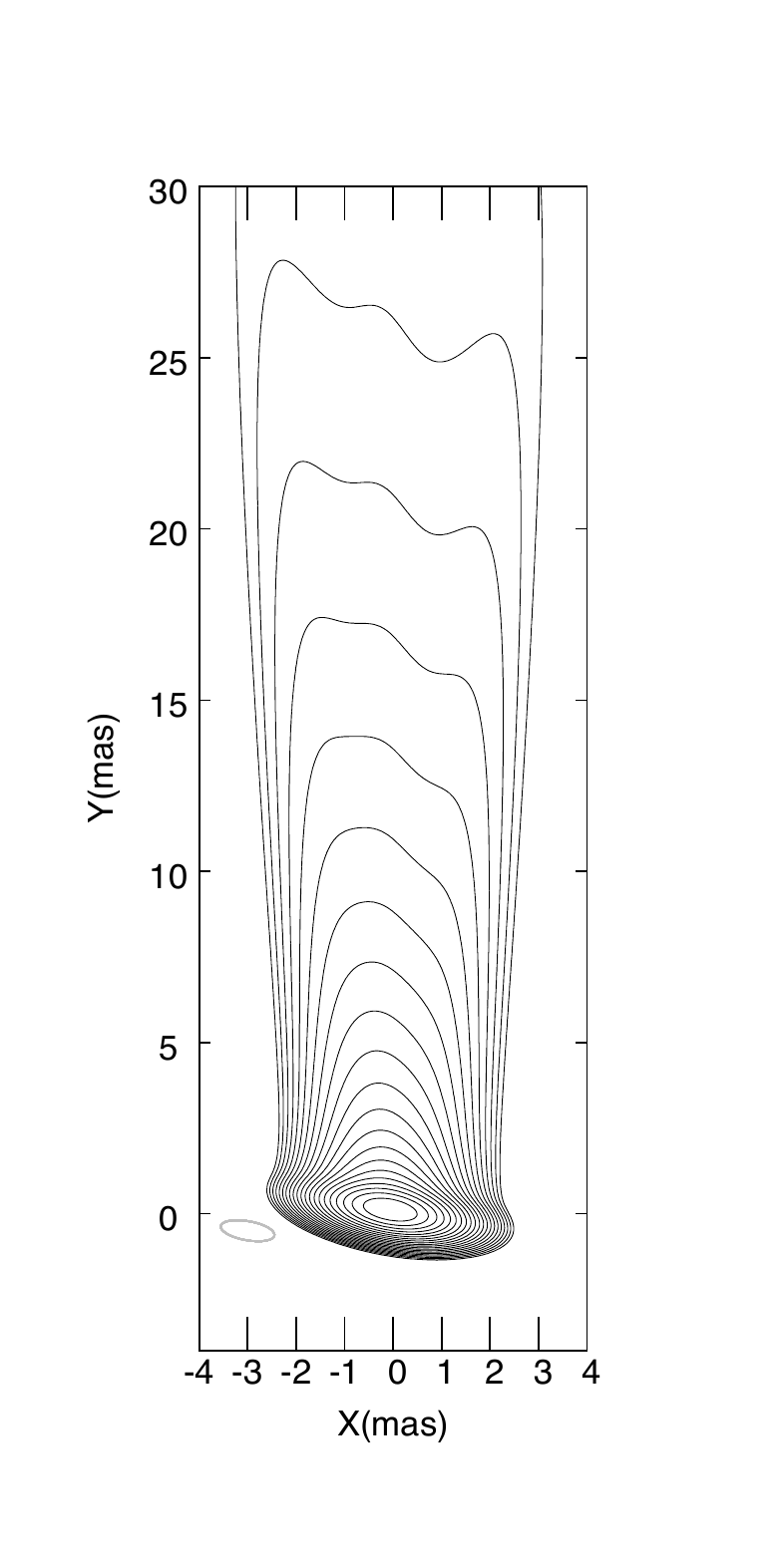}
  \caption{
    Left panel: The computed image with the computational resolution of $3R_S$.
    Right panel: The image convolved with the beam size ($1.14$ mas $\times$ $0.55$ mas), where the beam size is plotted as the gray circle at the bottom left corner.
    The intensity is normalized by each maximum value, and the contours represent the intensity at $2^{-k}$ ($k=1,2,3,...,27$ for the left panel and $k=1,2,3,...,21$ for the right panel).
    The number of the contour lines is not the same as in T18.
    Note that $1$ mas corresponds to $\approx 136R_{S}$.
  }
  \label{fig: result-images}
\end{figure*}

\begin{figure*}[htb!]
  \centering
  \includegraphics[clip,width=0.45\textwidth]{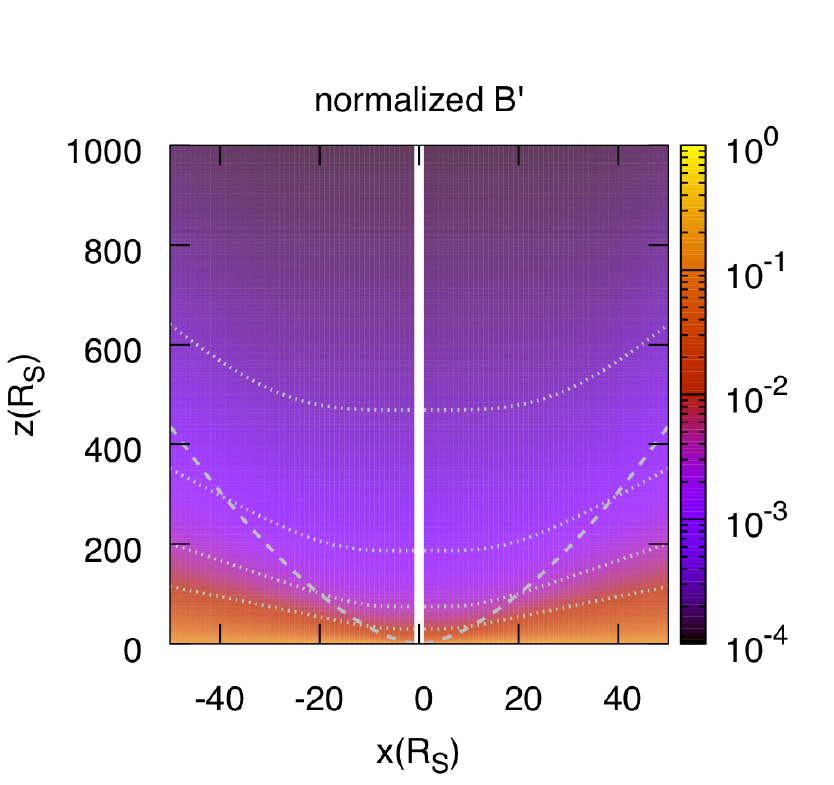}
  \includegraphics[clip,width=0.45\textwidth]{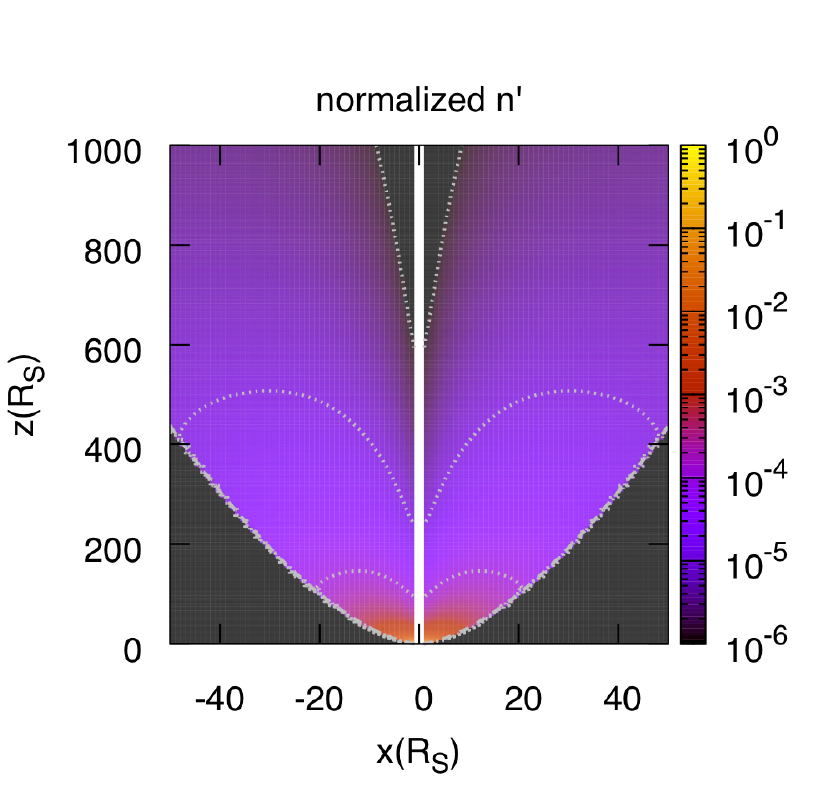}
  \includegraphics[clip,width=0.45\textwidth]{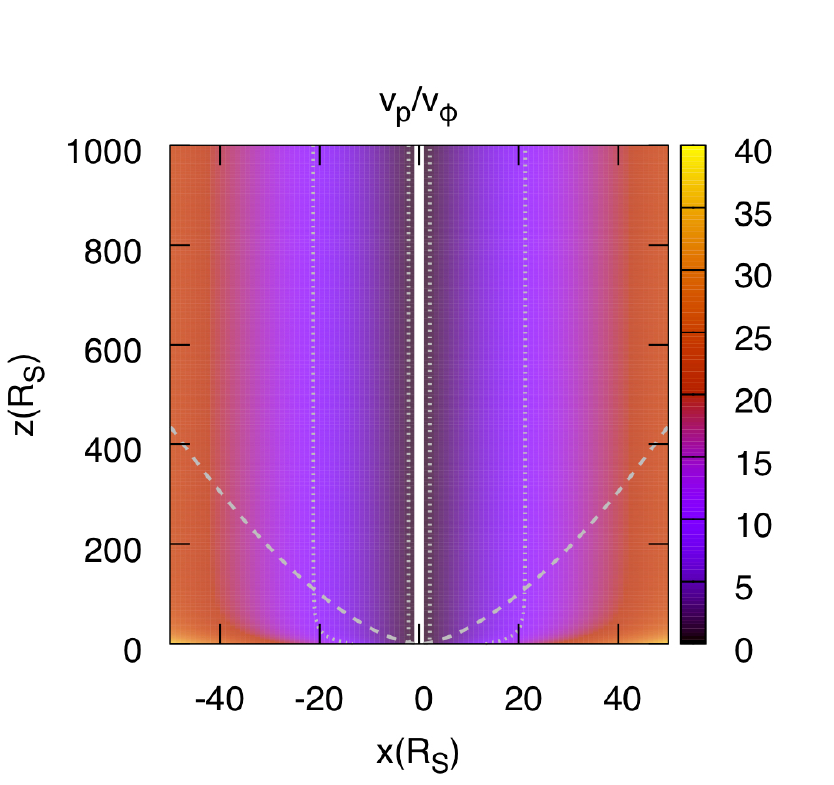}
  \includegraphics[clip,width=0.45\textwidth]{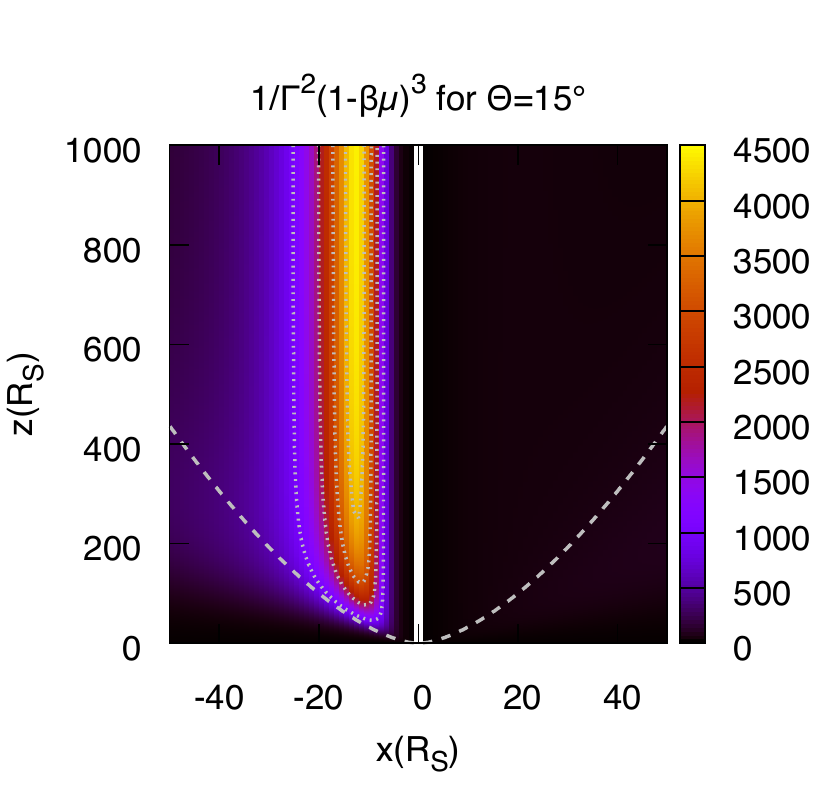}
  \caption{
    Spatial distribution of $B'$, $n'$, $v_p/v_{\phi}$, and $1/\Gamma^2(1-\beta\mu)^3$ for $\Theta = 15^\circ$ in the northern hemisphere.
    $B'$ and $n'$ are normalized by each maximum value.
    The white solid lines are the $z$ axis, and the gray dashed lines represent the field lines passing through the BH horizon at the equatorial plane.
    The dotted lines are the contours of $10^{-2}, 10^{-2.5}, 10^{-3},$ and $10^{-3.5}$ in the upper left panel, $10^{-4}, 10^{-5},$ and $10^{-6}$ in the upper right panel, $1$ and $10$ in the lower left panel, and $1000, 2000, 3000$, and $4000$ in the lower right panel.
    Note that $1R_{S}$ corresponds to $\approx 7.3 \mu$as.
  }
  \label{fig: density and B-field}
\end{figure*}

\subsection{Model Parameters} \label{subsec: model parameters}

Our model parameters are summarized in Table~\ref{tab:parameters}.
$a$, $\gamma'_{m}$, and $p$ are set to their fiducial values for which T18 obtained limb-brightened images.
While T18 used the same fiducial values for $\nu$, $\Theta$, and $M_{BH}$ as in \citet{Broderick2009} to compare their calculation results, we replace them with the values roughly consistent with recent observations of the M87 jet.
\citet{Nakamura2018} show that the width of the observed jet of M87 can be fitted by $z\propto R^{1.6}$ \citep[see also][]{Asada2012, Nakamura2013, Hada2013}.
Since Equation~(\ref{eq: Psi}) indicates that the shape of each field line obeys $z \propto R^{2/(2-\nu)}$ asymptotically, we adopt $\nu = 0.75$ assuming that the observed jet width reflects the magnetic field structure.
The BH mass and the viewing angle are set by $M_{BH}=6.2 \times 10^9 \, M_{\odot}$ \citep[][rescaled for $D = 16.7$ Mpc]{Gebhardt2011} and $ \Theta=15^{\circ}$ \citep{Wang2009}, respectively.
Then $1$ mas corresponds to $\approx 136R_{S}$ for $D = 16.7$ Mpc.
We use the beam size of 1.14 mas $\times$ 0.55 mas, for which a triple-ridge image of the M87 jet was obtained \citep{Hada2017}. Note that the beam size is larger than that in T18.

As for $R_p, \Delta,$ and $z_1$,
T18 showed the dependence of the image on $R_p$ with fixed values of $\Delta$ and $z_{1}$.
In the case of $R_{p}=0$, the jet image has one component around the axis, while in the case of $R_{p} > 0$ (e.g. $R_{p}=40\;R_{S}$, which is the fiducial value in their paper), limb-brightened images are obtained.
They focused on the region $\sim 1-4$ mas from the radio core, which is nearer than $\sim 10-30$ mas where the triple-ridge structure of the M87 jet is observed.
Even though we compute images with the same parameters as T18 for the far region, the triple-ridge structure cannot be obtained, as shown in Appendix~\ref{sec: appendixA}.
However, we found that the intensity map before the convolution for $R_{p}=0$ has a distinct bright component along the jet axis that may correspond to the observed inner-ridge (see Figure~\ref{fig: T18-Rp0}).
Then we fixed $R_{p}=0$, while setting large $\Delta$ to have a brighter jet edge, and found that the triple-ridge image can be obtained with $\Delta=2R_{S}$ and $z_{1}=5R_{S}$. We show the calculation results with these parameter values in the next section.

Note that $A$ and $n_0$ should be specified to obtain absolute intensity. T18 showed that reasonable values for them can lead to the observed level of intensity. However, the absolute intensity also depends on other uncertain parameters ($p$ and $\gamma'_{m}$) as well as the physical processes that we do not take into account in our current model such as the radiative cooling and reacceleration of electrons. In this paper, we treat $A$ and $n_0$ arbitrarily and focus on the relative intensity along the transverse direction of jets in order to simply illustrate our novel idea on the mechanism for the triple-ridge images.


\section{RESULTS} \label{sec: results}

In Section \ref{subsec: triple-ridge structure}, we show the resultant image of the triple-ridge structure, and figure out its physical origin.
In Section \ref{subsec: parameter dependence}, we demonstrate the parameter dependence of the image on the free parameters on the geometry of the jet $\Theta$ and $\nu$ and those on the electrons' spatial distribution $\Delta$ and $z_{1}$.


\subsection{Triple-ridge Structure} \label{subsec: triple-ridge structure}

Figure \ref{fig: result-images} shows the resultant image with the computational resolution (left panel) and the one convolved with the Gaussian beam (right panel).
The Gaussian beam size is represented as the gray circle in the bottom left corner of the right panel.
For both of the images, the intensity is normalized by each maximum value and the contours represent the intensity at $2^{-k}$ $(k=1,2,3,...)$.
In addition to the nearly symmetric two outer-ridges, the inner-ridge emerges in both images.
In the left panel we find that the inner-ridge extends along the jet axis and has an asymmetric shape. The right panel shows that the triple-ridge structure is highly smoothed by the convolution but still remains in $Y>15$ mas.
The counter jet appears in $Y<0$ in the left panel and overwhelmed by the bright core in the right panel.

The physical reason of this triple-ridge image can be explained by focusing on the dependence of $j_{\omega}$ on $n'$, $B'$ and the relativistic boosting factor $1/\Gamma^2(1-\beta\mu)^3$ (see Equations \ref{eq:jomega} and \ref{eq: Lorentz trans.}):
\begin{equation}
  j_{\omega} \propto \frac{n'B'^{(p+1)/2}}{\Gamma^2 (1-\beta\mu)^3}.
\end{equation}
We show the distribution of the three factors, as well as $v_p/v_\phi$, in Figure~\ref{fig: density and B-field}. Note that the distribution of $\sin\alpha'$ is not relevant to the image structure in this problem \citep[but see][for the problem on images of pulsar wind nebulae]{Shibata2003}. We confirmed that the shape of computed images with artificially setting $\sin\alpha' = 1$ do not change from those shown in Figure~\ref{fig: result-images}.

The upper left panel of Figure~\ref{fig: density and B-field} shows the profile of $B'$, which is roughly understood by writing down three magnetic field components (for $z > 0$) from Equations (\ref{eq: Bp}) and (\ref{eq: Bphi}),
\begin{eqnarray}
  B_{{\it r}} &=& Ar^{\nu-2} = \frac{\Psi}{R^2}(1+\cos\theta), \nonumber \\
  B_{\theta} &=& -\nu A r^{\nu-2} \frac{1-\cos\theta}{\sin\theta}= -\nu\frac{\Psi}{R^2}\sin\theta, \label{eq:Bcomponent}\\
  B_{\phi} &=& -\frac{2\Psi}{R^2} \frac{R\Omega_{\rm F}}{c^2}. \nonumber
\end{eqnarray}
For a given $R$, each component is roughly proportional to $\Psi$, so that it decreases for larger $z$.
The Lorentz transformation from the lab frame to the fluid frame does not significantly change the profile.
The upper right panel of Figure~\ref{fig: density and B-field} shows the density distribution.
At high $z$, the density around the axis becomes much smaller than the jet edge even though we set the density distribution concentrated around the axis at $z=z_{1}$.
This is because the distribution along the field line is derived from Equation~(\ref{eq:nb2const}) and the magnetic field is stronger around the jet edge.
Therefore, one can see that the bright outer-ridges are produced by the strong magnetic field and high density around the jet edge.

The magnetic field is weaker and the electron density is lower around the axis, but the strong relativistic beaming makes the bright inner-ridge.
As seen in Equation~(\ref{eq:Bcomponent}) the magnetic field is poloidal dominant at $R \ll c/\Omega_{F}$ and toroidal dominant at $R \gg c/\Omega_{F}$ ($c/\Omega_{F} \simeq 2.13\;R_{S}$). Correspondingly, the fluid velocity is toroidal dominant at $R \ll c/\Omega_{F}$ and poloidal dominant at $R \gg c/\Omega_{F}$ (see the lower left panel of Figure~\ref{fig: density and B-field}).
In between, there is a region where the fluid velocity is almost parallel to the line of sight, i.e., $\mu \approx 1$.
This enhances the factor $1/\Gamma^2(1-\beta\mu)^3$, as shown in the lower right panel of Figure~\ref{fig: density and B-field}.
Although the magnetic field, electron density, and the Lorentz factor are small near the jet axis, the beaming effect produces the bright sharp inner-ridge.

The transverse width of the relativistically beamed area, $\sim 10R_{S} \sim 0.1$ mas, shown in the bottom right panel of Figure~\ref{fig: density and B-field} corresponds to the width of the inner-ridge of the left panel of Figure~\ref{fig: result-images}.
We expect that observations with better resolution can provide the narrower inner-ridge.
Besides, the inner-ridge appears only in one side of the jet axis.
Which side the inner-ridge appears depends on the sign of $\Omega_{F}$ and tells us the direction of the BH rotation.


\begin{figure*}
  \centering
	\includegraphics[clip,width=0.45\textwidth]{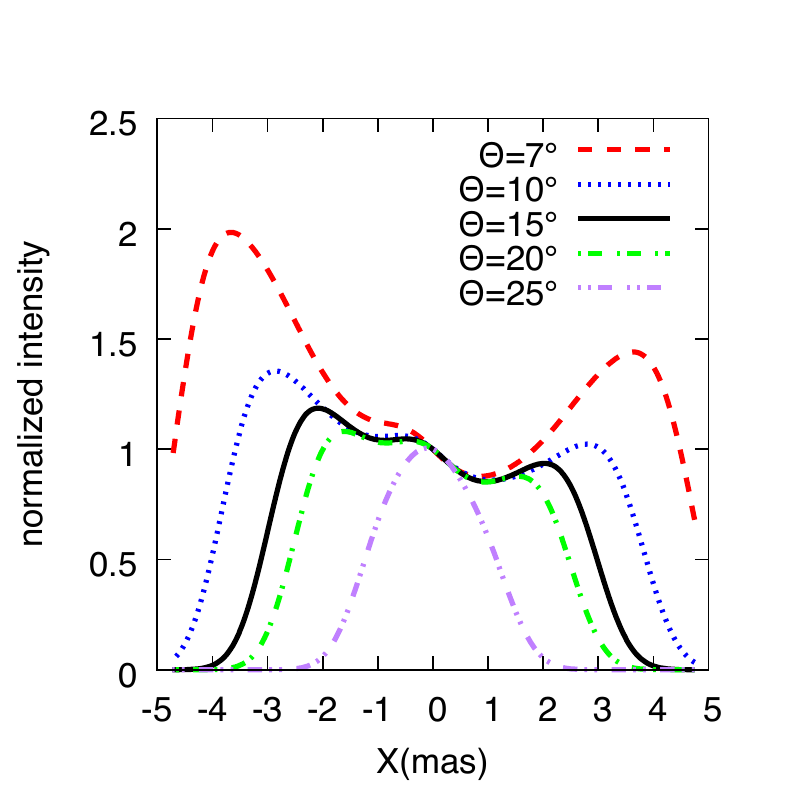}
	\includegraphics[clip,width=0.45\textwidth]{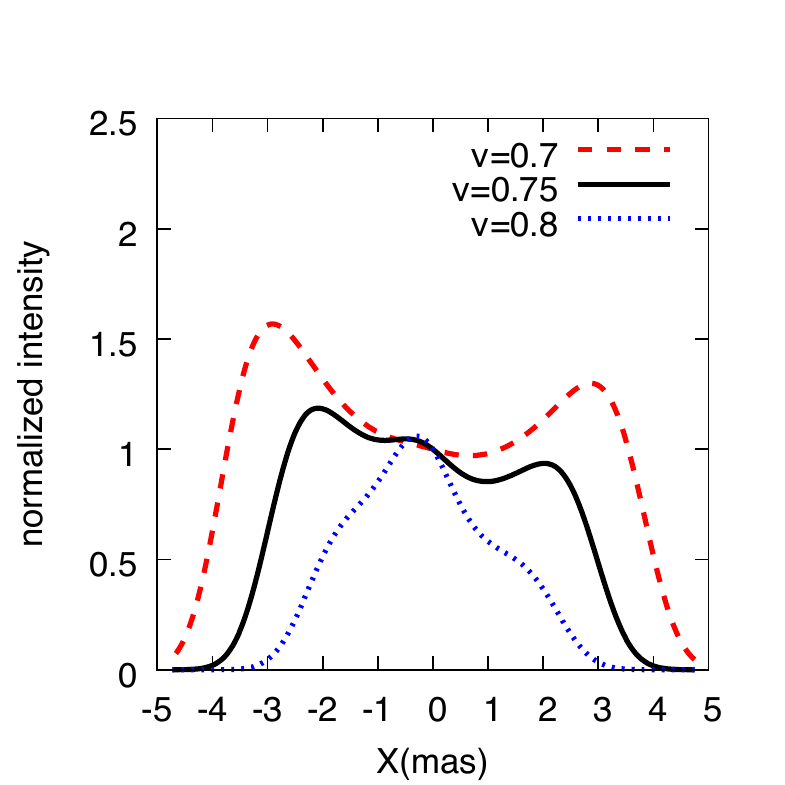}
	\includegraphics[clip,width=0.45\textwidth]{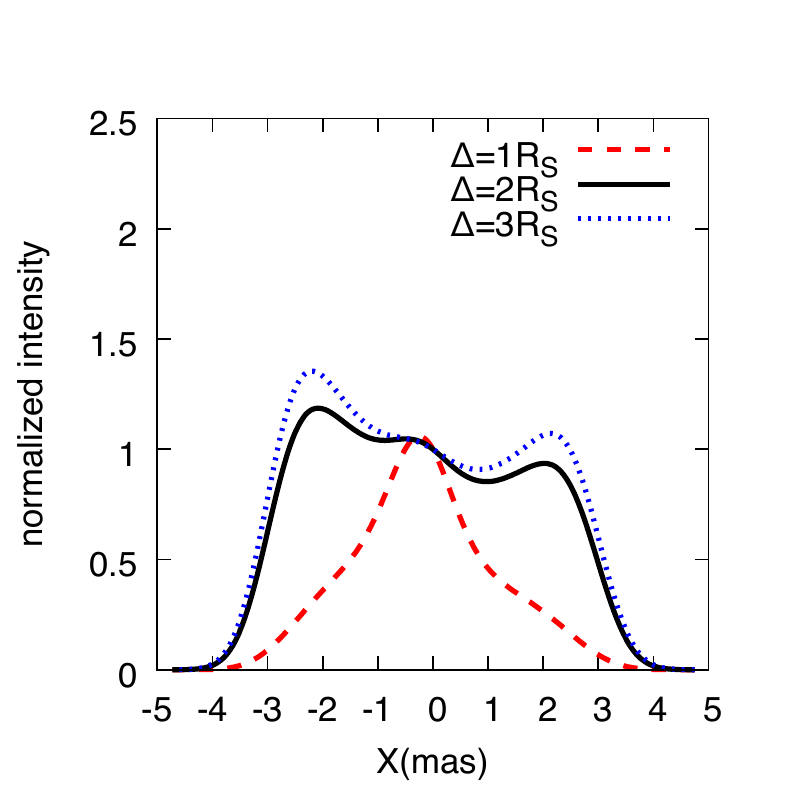}
	\includegraphics[clip,width=0.45\textwidth]{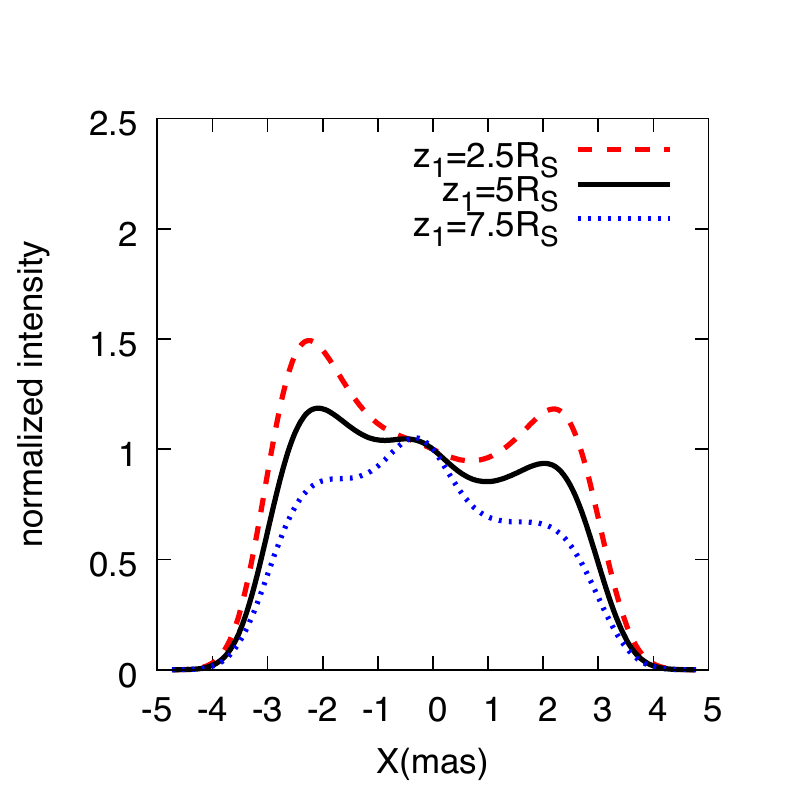}
	\caption{
	  Dependence of the transverse intensity profile at $Y=25$ mas on $\Theta$ (upper left), $\nu$ (upper right), $\Delta$ (lower left), and $z_{1}$ (lower right).
	  The solid line in each panel is identical to the case of Figure~\ref{fig: result-images}.
    The dashed, solid, dotted, dashed-dotted, and dashed double-dotted lines represent the results calculated with changing the parameter values from the case of the solid line as shown in each panel.
	  The vertical axis is normalized by the value at $X=0$ of each line.
	}
  \label{fig: dependence}
\end{figure*}

\subsection{Parameter Dependence} \label{subsec: parameter dependence}
We calculate the images with different $\Theta, \nu, \Delta$ and $z_{1}$ to examine the parameter dependence on the image structure.
For illustration, we take $Y=25$ mas, for which we plot the resultant transverse intensity profiles in Figure~\ref{fig: dependence}. Note that we normalize the intensity by the value at $X=0$ of each line in Figure~\ref{fig: dependence}.
The upper left panel represents the intensity profiles in the case of $\Theta =$ $7^{\circ},$ $10^{\circ},$ $15^{\circ},$ $20^{\circ},$ and $75^{\circ}$ with the other parameters unchanged.
For smaller $\Theta$, the relation $z=Y/\sin\Theta$ for a fixed value of $Y$ means that the line of sight passes the points farther from the BH, and then the jet width appears to be larger.
For larger $\Theta$, the outer-ridges are debeamed because the line of sight and the beaming cone of the outer-ridges are misaligned, while the inner-ridge remains.

The upper right panel of Figure~\ref{fig: dependence} shows the intensity profiles in the case of $\nu = 0.7, 0.75,$ and $0.8$ with the other parameters unchanged. When $\nu$ increases, the field lines and density concentrate to the axis.
This leads to a candle-flame-like shaped image.
On the other hand, when $\nu$ decreases, the field lines and density distribution expand. This results in the bright outer-ridges, which dominate the inner-ridge component.

As for the dependence on the parameters related to the density distribution, we examine only the dependence on $\Delta$ and $z_{1}$ because the dependence on $R_{p}$ is already discussed in T18.
As $\Delta$ increases or $z_{1}$ decreases, the density in the jet edge increases and the outer-ridges become brighter as shown in the lower panels in Figure~\ref{fig: dependence}.
The positions $X$ of the outer-ridge peaks do not change between the cases of $\Delta=2 R_{S}$ and $\Delta=3 R_{S}$, and between the cases of $z_{1}=2.5 R_{S}$ and $z_{1}=5 R_{S}$ because the jet width is limited by the outermost magnetic field line threading the BH.
Although our setup of the electron density distribution is a toy model, this analysis indicates that the observed image structure can strongly constrain the spatial electron density distribution when $\Theta$ and $\nu$ are estimated by other observational information such as the blob pattern speed, the brightness ratio between the approaching and counter jets, and the width profile of the jet.

\begin{figure*}[htb!]
  \centering
  \includegraphics[clip,width=0.75\textwidth]{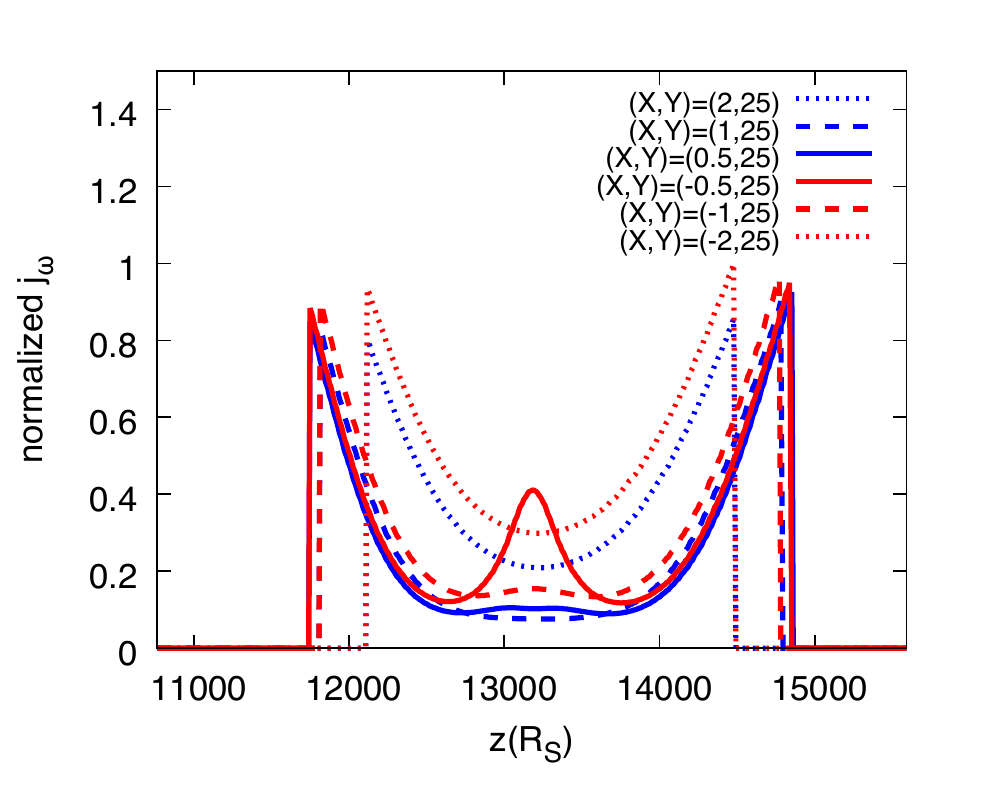}
  \caption{Distribution of $j_\omega$ along the line of sight of (X, Y) = ($\pm$2 mas, 25 mas) (dotted lines), ($\pm$1 mas, 25 mas) (dashed lines), ($\pm$0.5 mas, 25 mas) (solid lines). The red (blue) lines are for negative (positive) X cases. The horizontal axis represents the height $z$ at the point on the line of sight. All the values of $j_\omega$ are normalized by the maximum value for $(X, Y)=$($-2$ mas, 25 mas).}
  \label{fig: los}
\end{figure*}


\section{DISCUSSION} \label{sec: discussion}

\subsection{Valleys between Ridges} \label{subsec: Valleys between Ridges}
As seen in Figures~\ref{fig: result-images} and \ref{fig: dependence} the inner-ridge and outer-ridges produced by our model are not so clearly separate as reported in \citet{Hada2017}.
This is because the jet edge have the sheath-like three dimensional structure and the emission from the sheath enhances the brightness of the valleys between the ridges.

For more details, we show $j_{\omega}$ distributions along lines of sight passing $(X, Y)$ = ($\pm$0.5 mas, 25 mas), ($\pm$1 mas, 25 mas), and ($\pm$2 mas, 25 mas) as functions of $z$ in Figure~\ref{fig: los} to examine which parts of the jet contribute to the transverse intensity profile $I_\omega (X, Y=25\;{\rm mas})$ (the black lines in Figure~\ref{fig: dependence}).
The jet emission is composed of the sheath-like jet edge component and the beamed inner-ridge component.
For the line of sight of $(X, Y)=$ ($-0.5$ mas, 25 mas), the inner-ridge component appears in addition to the jet edge component, while for the line of sight of $(X, Y)=$ (0.5 mas, 25 mas), the emission around the axis is debeamed.
This leads to the asymmetry of the inner-ridge in our computed image, which we have pointed out for the left panel of Figure~\ref{fig: result-images} in Section~\ref{sec: results}.

The lines of sight passing $(X,Y)$ = ($\pm$1 mas, 25 mas) and ($\pm$2 mas, 25 mas), for which the valleys and outer-ridges are seen, respectively (see Figures~\ref{fig: result-images} and \ref{fig: dependence}), penetrate only the jet edge.
Interestingly, $j_{\omega}$ at the rear part of jet edge $z \lesssim 1.25\times 10^4\;R_{S}$ is comparable to that at the front part of jet edge $z \gtrsim 1.4\times 10^4\;R_{S}$ for each of these sight lines.
That is because the fluid motion at the rear part directs away from the line of sight and then the emission is debeamed, but $n'$ and $B'$ are larger there than those at the front part with higher $z$.
Figure~\ref{fig: los} implies that the intensities $I_\omega = \int j_\omega dZ$ for $(X,Y)$ = ($\pm$1 mas, 25 mas) and ($\pm$2 mas, 25 mas) should be comparable, which means that the valleys are not deep.

The observed deep valleys may indicate more complex structure of the jet.
For example, if the non-thermal electrons distributed separately at the spine and a thin layer of the jet edge, then the brightness for $(X,Y)$=($\pm$1 mas, 25 mas) would become lower.
A thin layer with dense non-thermal electrons would produce bright outer-ridges with deep valleys.\footnote{From a simple geometric consideration for the jet with a thin layer of the width $\Delta R$ viewed at $\Theta=90^{\circ}$, one can see that the ratio of the path lengths along the lines of sight across the thin layer for the outer-ridge and valley scales as $\propto \sqrt{\Delta R}$.}
We also consider that non-axisymmetric jet may lead to deep valleys.
For example, if only the rear part of the jet edge at $-1\;{\rm mas} \lesssim X \lesssim 1\;{\rm mas}$ is intrinsically dim, the intensities at the valleys become about half with keeping the outer-ridges bright.


\subsection{Bulk Lorentz Factor at the Far Zone} \label{subsec: Bulk Lorentz Factor at the Far Zone}

\begin{figure}
	\centering
	\includegraphics{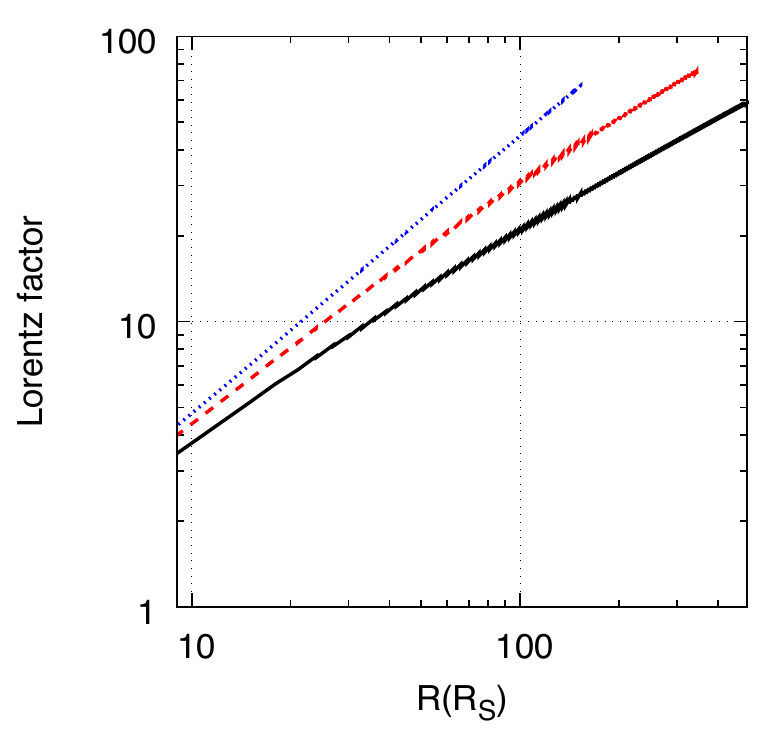}
	\includegraphics{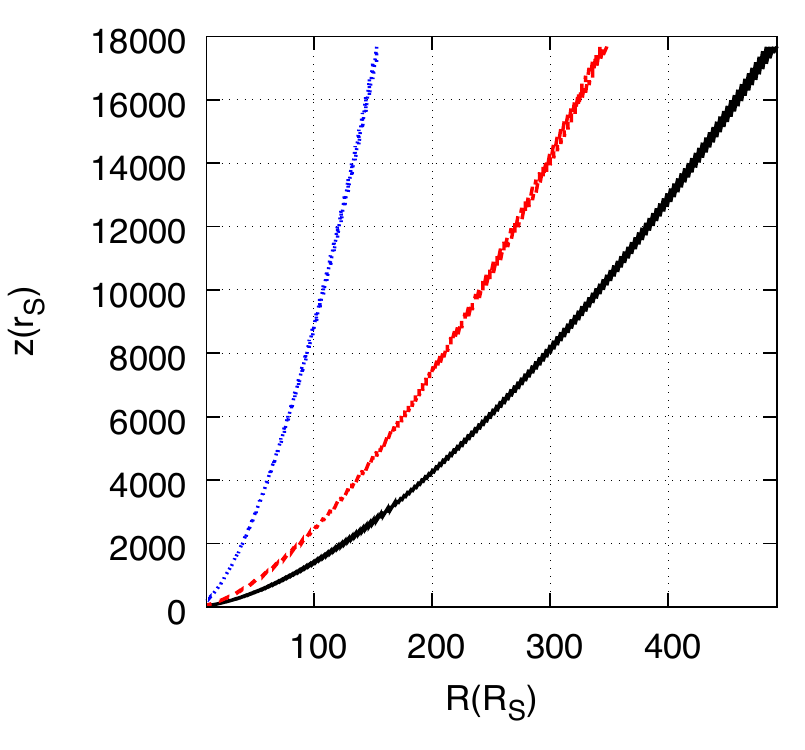}
	\caption{ Upper panel: Lorentz factor profiles along the outermost field line of $\Psi = \Psi(r_{+},\pi/2)$ (black solid line), $\Psi = \Psi(r_{+},\pi/2) /2$ (red dashed line) and $\Psi = \Psi(r_{+},\pi/2) /10$ (blue dotted line) in the model of Figure~\ref{fig: result-images}. Lower panel: The shape of the field lines of the upper panel. The upper panel is plotted in log scale and the lower panel is in linear scale. }
	\label{fig: Lorentz factor}
\end{figure}

We assume that the force-free approximation is valid even at the far zone from the BH. Then the bulk Lorentz factor $\Gamma$ keeps increasing. Figure~\ref{fig: Lorentz factor} represents the Lorentz factor profiles along the magnetic field lines which satisfy $\Psi = \Psi (r_{+}, \pi/2)$, $\Psi = \Psi (r_{+}, \pi/2) /2$, and $\Psi = \Psi (r_{+}, \pi/2) /10$ in the model of Figure~\ref{fig: result-images}. These obey analytical relation
$1/\Gamma^2 = 1/\Gamma_1^2 + 1/\Gamma_2^2$, where $\Gamma_1^2 = B^2/B_{p}^2 \propto R^2$ and $\Gamma_2^2 = B^2/(B_{\phi}^2 - E^2) \propto (R_{c}/R)$ at the far region $R\Omega_F \gg c$, where
$R_{c}$ is the curvature radius of the poloidal field line.
This asymptotic relation is shown in \citet{Tchekhovskoy2008} for the force-free case\footnote{The asymptotic relation (47) of the Lorentz factor in T18 is not complete but valid only for the second acceleration regime. In the first acceleration regime, $\Gamma$ is given as $\Gamma \sim \sqrt{1 + (\frac{R\Omega}{cg(\nu,\theta)})^2}$, which is proportinal to $R$ when the second term dominates.} and in \citet{Komissarov2009} and \citet{Lyubarsky2009} for the Poynting-dominated cold ideal MHD case.

The Lorentz factor increases up to $\sim 80$ in the computated area along each field line shown in Figure~\ref{fig: Lorentz factor}. This is much higher than the Lorentz factors deduced from blob motions observed in the VLBI observations \citep[][and references there in]{Mertens2016, Hada2017}, $\Gamma \lesssim 10$, although it is possible that faster blobs are not identified due to the apparent decrease of number of fast blobs \citep{Komissarov1997} and low time-resolution of the current monitoring \citep{Nakamura2018}.
If the Lorentz factors of currently identified blob motions are similar to those of the steady flows in jets, we require MHD models with Poynting to kinetic energy flux ratio (i.e., $\sigma$ parameter) at the jet base as low as $\sim 10$, which gives lower saturation Lorentz factors.

\subsection{Inner-ridge Property} \label{subsec: inner-ridge property}

The inner-ridge of our computed image arises since there is a region where the direction of the velocity is parallel to the line of sight between the toroidal dominant region ($v_{\phi} \gg v_{p}$, $R\Omega_F/c \ll 1$) and poloidal dominant region ($v_{p} \gg v_{\phi}$, $R\Omega_F/c \gg 1$), as explained in Section~\ref{subsec: triple-ridge structure}. 
We have demonstrated this property by setting the fluid velocity to be the drift velocity (Equation~\ref{eq: drift velocity}).
 The cold ideal MHD velocity also has the same property as the drift velocity, i.e., it is toroidal dominant at $R\Omega_{F}/c \ll 1$ while poloidal dominant at $R\Omega_{F}/c \gg 1$ \citep[][T18]{Beskin2010, Toma2013,Pu2015}. 
 
\begin{figure}
	\centering
	\includegraphics{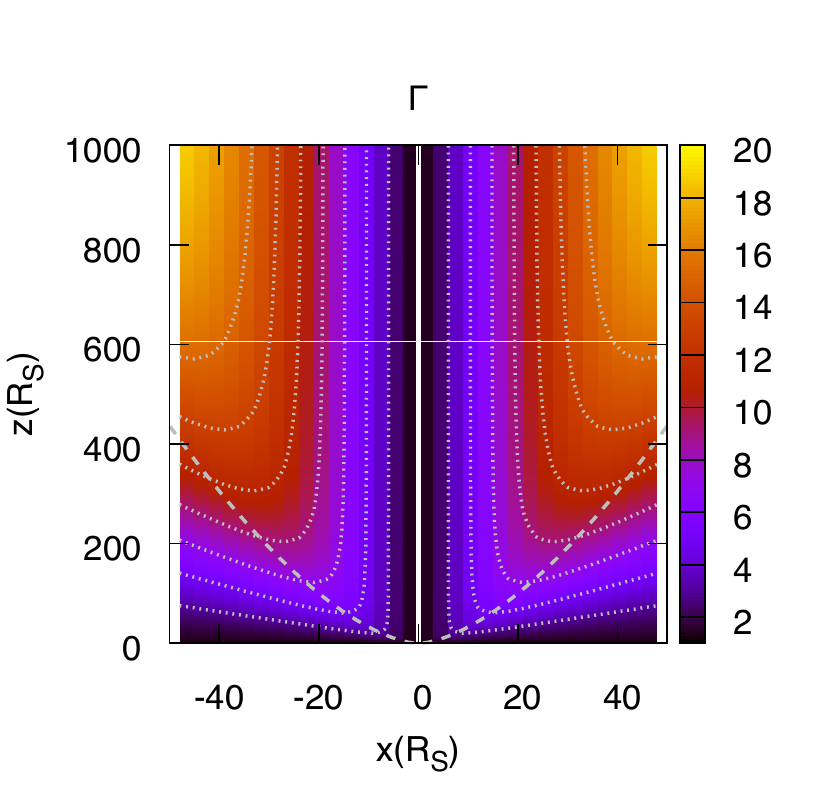}
	\caption{ Spatial distribution of $\Gamma$. The dotted lines are the contours of $3, 5, 7, 9, 11, 13, 15$. The white solid line and the gray dashed line are the same for Figure~\ref{fig: density and B-field}. }
	\label{fig: gamma }
\end{figure}

However, if the jet particles are injected along the magnetic field lines with the large Lorentz factor, $\Gamma_{in}$, at the jet base, the velocity structure will change, i.e., the velocity at the region $R \Omega_F/c \ll 1$ where $B_p \gg B_{\phi}$ will become dominated by the poloidal component \citep{Beskin1997,Beskin2000}.
 We plot the Lorentz factor close to the axis for our fiducial model in Figure~\ref{fig: gamma }, which shows that $5 \lesssim \Gamma \lesssim 9$ around the region that produces the inner-ridge ($c/\Omega_F \lesssim 10R_S \lesssim R \lesssim 20R_S$, cf. the bottom-right panel in Figure~\ref{fig: density and B-field}). 
If $\Gamma_{in}$ is smaller than $5$, the velocity structure at $R \gtrsim 10 R_S$ will not be changed, so that the triple-ridge structure will be still observed.
For a smaller viewing angle, the inner-ridge radius is larger (as seen in the upper left panel of Figure~\ref{fig: dependence}) and the Lorentz factor at that radius is larger, so that the triple-ridge structure will still appear in cases of even larger $\Gamma_{in}$. 
We note that the large $\Gamma_{in}$ will also change the field configuration because of the particle inertia.
These effects have not been taken into account either in GRMHD simulations. We need a more sophisticated model to consider them in detail.

The lower right panel of Figure~\ref{fig: density and B-field} shows that the inner-ridge radius does not depend on $z$.
In contrast, \citet{Asada2016} indicates that the inner-ridge width of the M87 jet varies as a function of the distance from the BH.
They showed that further from the BH, the inner-ridge becomes wider at 5 GHz, while it becomes narrower at 1.6 GHz.
This might be caused by the synchrotron cooling as mentioned in \citet{Asada2016} or the time variation of the jet, which are not included in our model.

\section{SUMMARY} \label{sec: summary}
We have examined a steady axisymmetric force-free model of a jet driven by BH, in which the electromagnetic structure is set to be consistent with GRMHD simulation results, and shown that the triple-ridge structure of a relativistic jet can be produced by the model with a simple Gaussian distribution of emitting electrons at the jet base ($z=z_1$).
We have found that the fluid drift velocity associated with such field structure produces the inner-ridge by the relativistic beaming effect, and the magnetic field strength and electron number density are higher nearer the jet edge, which produce the outer-ridges.
Thus we argue that the observed triple-ridge image does not directly indicate the requirement of the two jet launching processes working simultaneously such as the BZ process for the rotating BH plus the BP process for the accretion disk.
This argument appears to be consistent with the finding of T18 that the jet from the accretion disk produces highly asymmetric images unlike the observed limb-brightened images of the M87 jet and with the GRMHD simulation result of \citet{Nakamura2018} that the outermost parabolic streamline of the jet driven by BZ process overlaps the edge of the observed M87 jet.

We also have found that the jet image is very sensitive to the height $z_1$ of the base of electron spatial distribution and the width of its distribution $\Delta$, as well as to the geometric parameters of jet $\Theta$ and $\nu$. This means that the characteristic jet images as observed with recent sensitive VLBI observations at 43 and 86 GHz can strongly constrain the spatial distribution of injected electrons near the BH, when $\Theta$ and $\nu$ are estimated by other observational information such as the brightness ratio between the approaching and counter jets and the width profile of the jet. Such studies must be complementary to those directly investigating physics closely around the BH with the upcoming EHT data \citep{Doeleman2012,Akiyama2017}.

Our model does not reproduce the sharpness of the observed ridges and the width variation of the inner-ridge as a function of $z$ in the M87 jet. They may be caused by distinct production of the non-thermal electrons at the spine and sheath, non-axisymmetry of the jet, temporal variation of the jet, and synchrotron cooling and/or reacceleration of electrons. To include such effects in the jet model is required as separate work.


\acknowledgments
We thank the participants in the Mizusawa Project Meetings in 2016, 2017, and 2018 for fruitful discussions on relativistic jets.
Numerical calculations were performed on Draco, a computer cluster of the Frontier Research Institute for Interdisciplinary Sciences in Tohoku University.
TO acknowledges financial support from the Graduate Program on Physics for the Universe of Tohoku University.
This work is partly supported by JSPS Grants-in-Aid for Scientific Research 17H06362 (K.~Takahashi), 18H01245 (K.~Toma), and also by a JST grant Building of Consortia for the Development of Human Resources in Science and Technology (K.~Toma).


\appendix
\section{T18 Model at the Far Zone} \label{sec: appendixA}

T18 showed the limb-brightened images at $\sim 1-4$ mas from the core with parameter values $M_{BH}=3.4 \times 10^9 \, M_{\odot},$ $a=0.998$, $\Omega_{F}$ = $0.5 \Omega_{H}$, $\Theta=25^{\circ}$, $\nu = 1$, $z_1 = 5 R_{S}, \Delta = 5 R_{S}$, and the beam size $0.43$ mas $\times 0.21$ mas \citep{Walker2008}.
We show the resultant synchrotron images with these parameter values up to 30 mas, which do not exhibit the triple-ridge structure.

\begin{figure}[htb!]
  \centering
  \includegraphics[clip,width=0.45\textwidth]{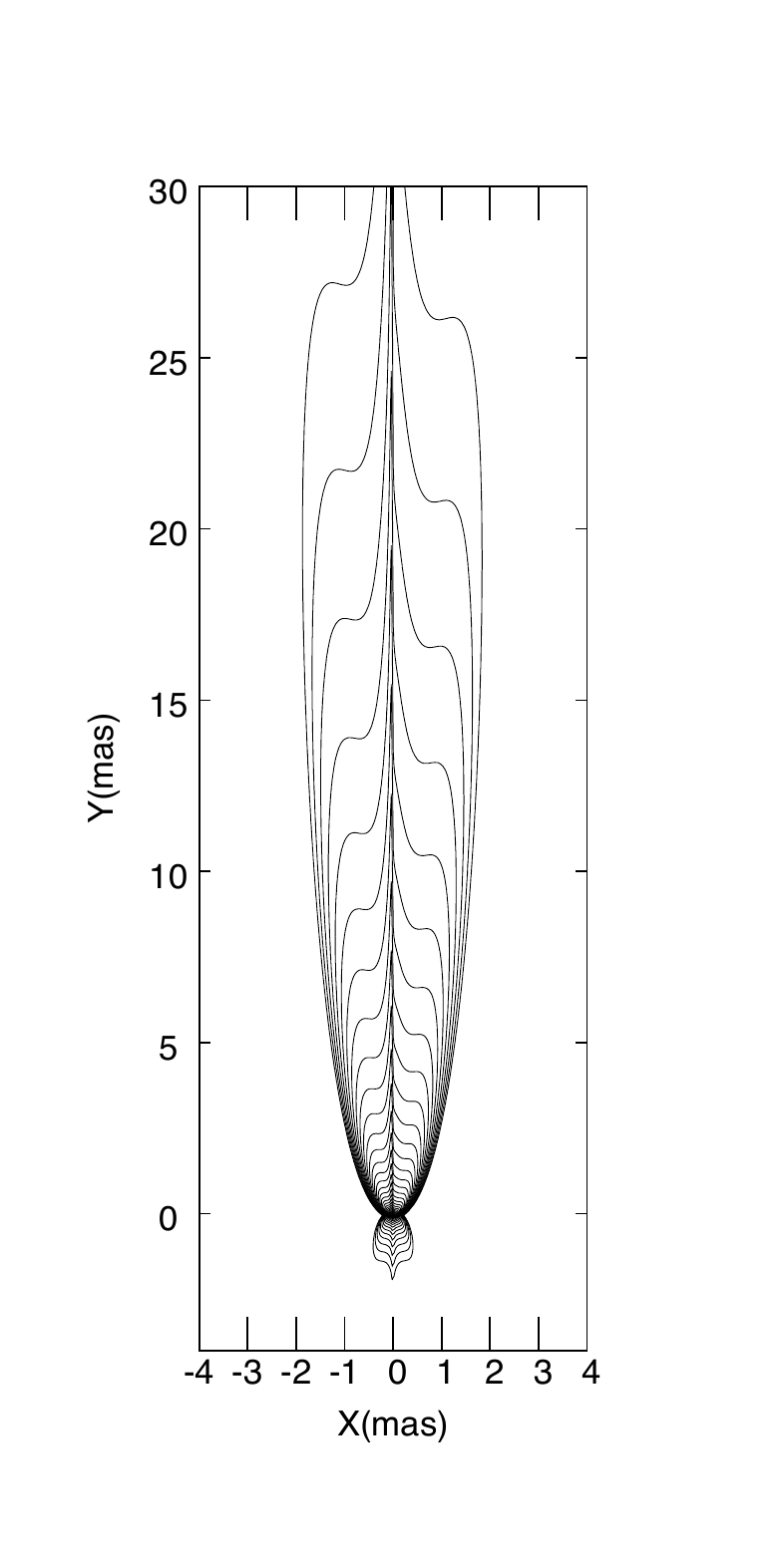}
  \includegraphics[clip,width=0.45\textwidth]{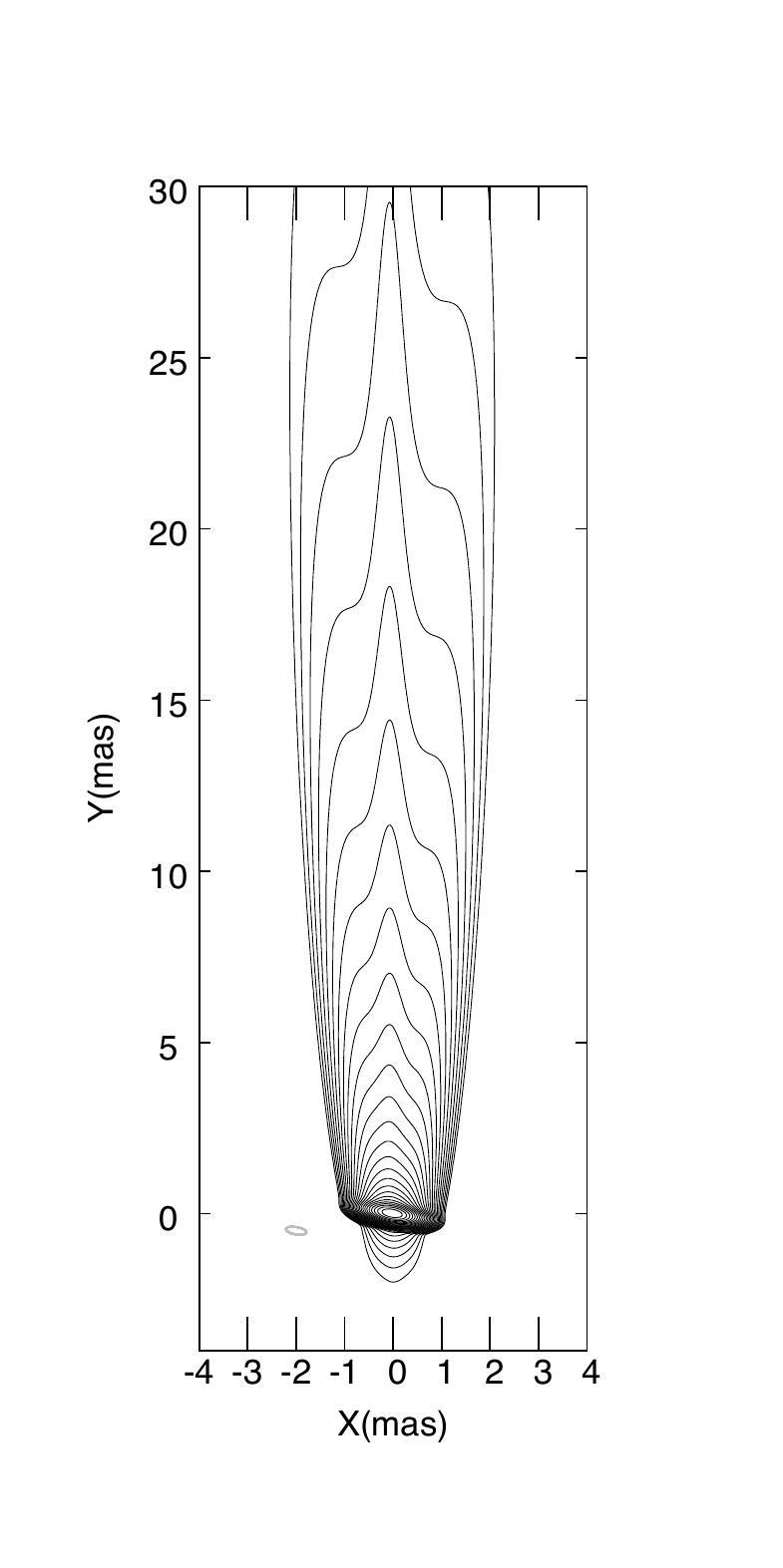}
  \caption{The synchrotron images with the same parameters as the case of $R_{p}=0$ in T18. Left: Image with the computational resolution of $5R_{S}$. Right: Image convolved with the beam size (0.43 mas $\times$ 0.21 mas). The contours represent the normalized intensity at $2^{-k}$ ($k=1,2,3,...,30$ for the left panel and $k=1,2,3,...,26$ for the right panel).
  }
  \label{fig: T18-Rp0}
\end{figure}

Figure \ref{fig: T18-Rp0} shows the result for the case of $R_{p}=0$. The jet image does not have the triple-ridge structure even at $Y > 4$ mas, but a narrow bright component near the axis is remarkable, which corresponds to the inner-ridge from our viewpoint.

\begin{figure}[htb!]
  \centering
  \includegraphics[clip,width=0.45\textwidth]{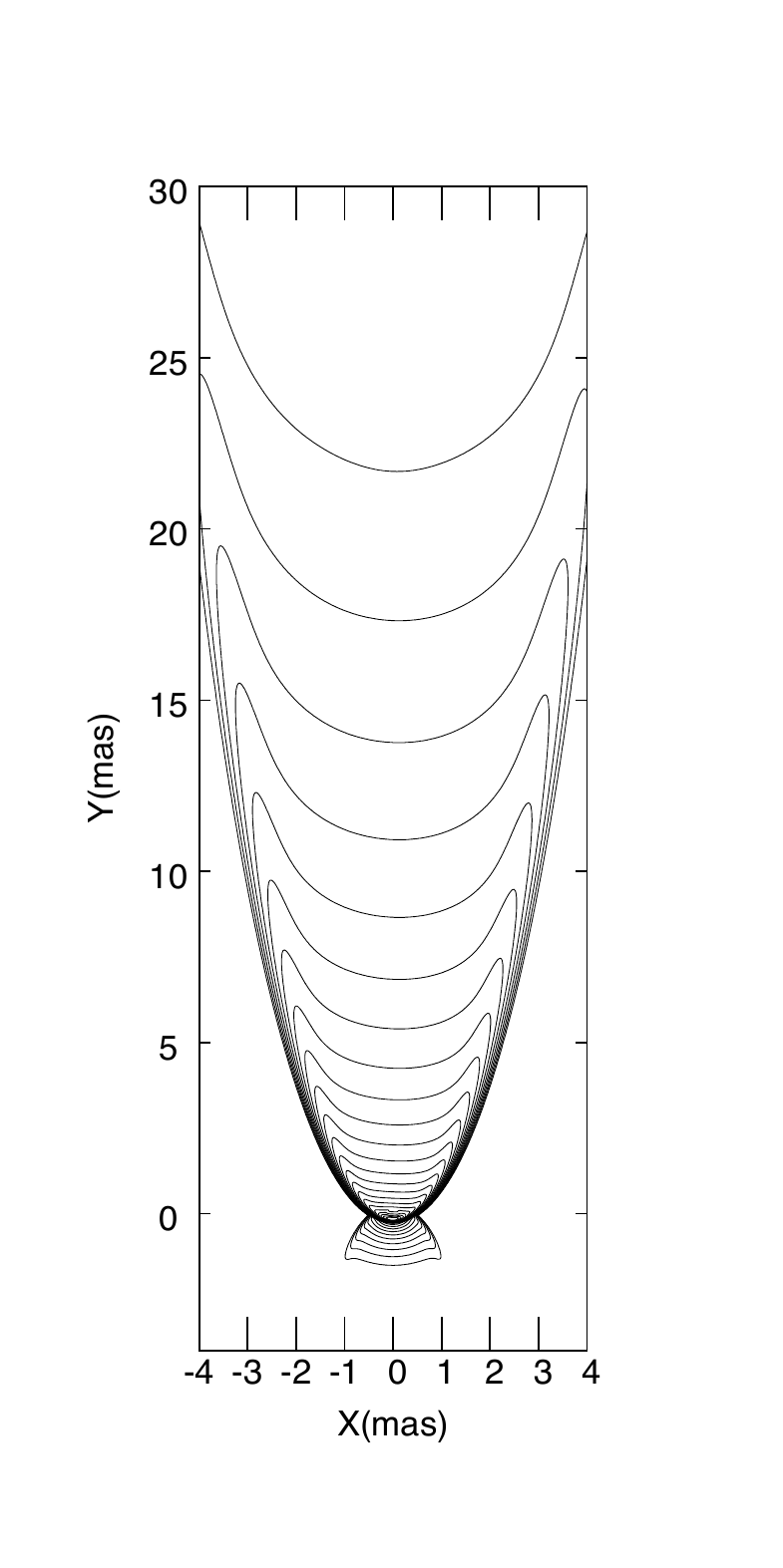}
  \includegraphics[clip,width=0.45\textwidth]{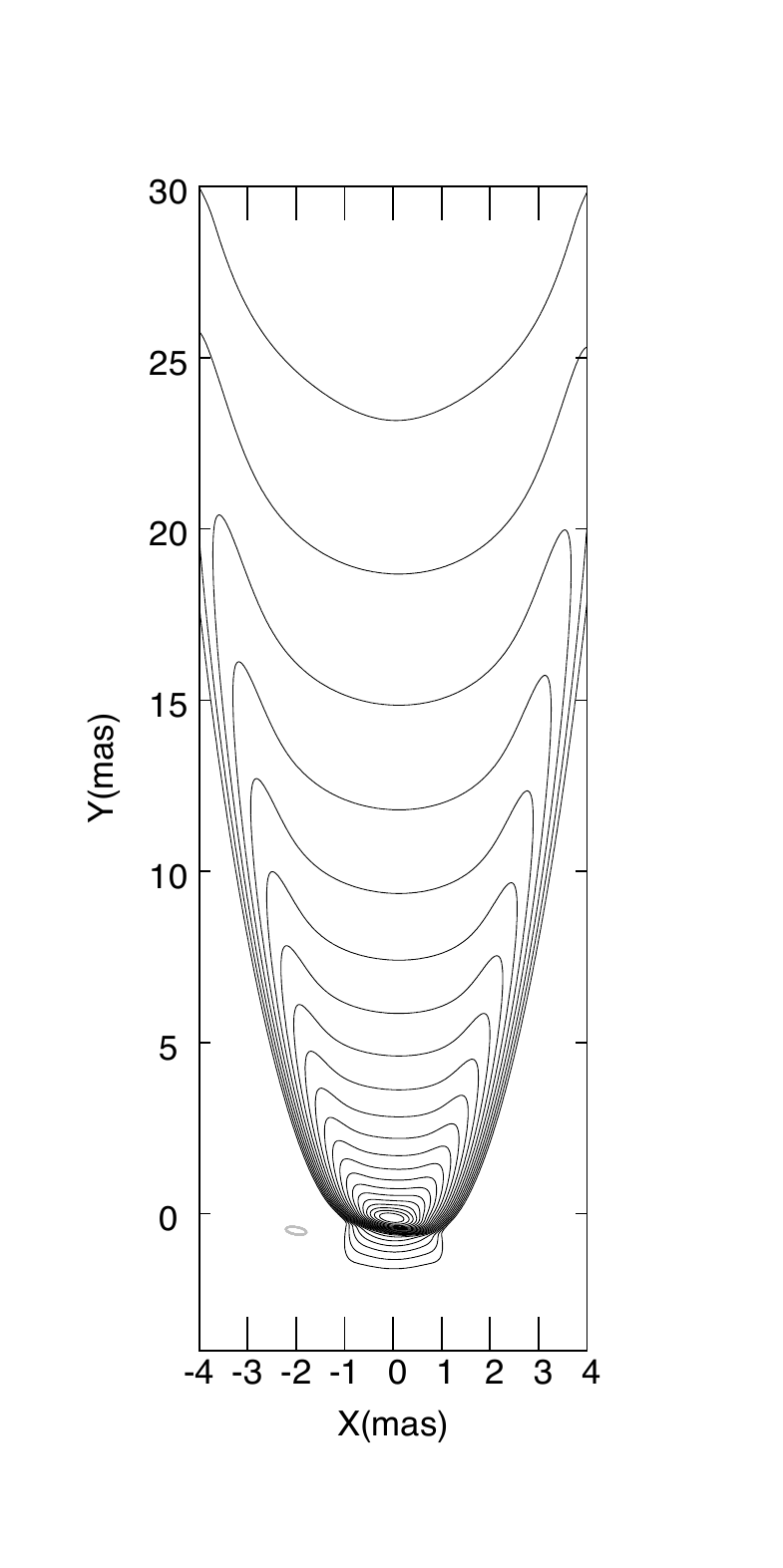}
  \caption{The synchrotron images with the same parameters as the case of $R_{p}=40 R_{S}$ in T18. Left: Image with the computational resolution of $5R_{S}$. Right: Image convolved with the beam size (0.43 mas $\times$ 0.21 mas). The contours represent the normalized intensity at $2^{-k}$ ($k=1,2,3,...,23$ for the left panel and $k=1,2,3,...,21$ for the right panel).
  }
  \label{fig: T18-Rp40}
\end{figure}

Figure \ref{fig: T18-Rp40} shows the result for the case of $R_{p}=40R_{S}$. The limb-brightened structure is obtained, and no clear inner-ridge was found in this case. The ring-like distribution of electrons does not produce high emissivity around the axis.

\bibliography{./draft}

\end{document}